\documentclass[a4paper]{jpsj2}

\def\runauthor{Shinya {\sc Saka} and Hiroshi {\sc Takano}}
\title{%
	Relaxation of a Single Knotted Ring Polymer
}
\author{%
	Shinya {\sc Saka}\thanks{E-mail address: ssaka@rk.phys.keio.ac.jp}
	and
	Hiroshi {\sc Takano}
}
\inst{%
	Faculty of Science and Technology,
	Keio University, Yokohama 223-8522, Japan
}
\abst{%
The relaxation of a single knotted ring polymer
is studied by Brownian dynamics simulations. 
The relaxation rate $\lambda_q$ for the wave number $q$ is estimated
by the least square fit of
the equilibrium time-displaced correlation function
$
	\hat{C}_q(t) = N^{-1} \sum_i \sum_j
	\frac{1}{3} \langle  \mib{R}_i (t) \cdot \mib{R}_j (0) \rangle
	\exp [{\rm i}{2\pi}q(j-i)/N]
$
to a double exponential decay at long times. 
Here,
$N$ is the number of segments of a ring polymer and
$\mib{R}_i$ denotes
the position of the $i$th segment
relative to the center of mass of the polymer.
The relaxation rate distribution
of a single ring polymer with the trivial knot
appears to behave as
$ \lambda_q \simeq A(1/N)^x $ for $q=1$ and
$ \lambda_q \simeq A'(q/N)^{x'} $ for $q>1$,
where
$x \simeq 2.10$, $x' \simeq 2.17$ and $A < A'$.
These exponents are similar to that found for a linear polymer chain. 
The topological effect appears as the separation of 
the power law dependences for $q=1$ and $q >1$,
which does not appear for a linear polymer chain. 
In the case of the trefoil knot,
the relaxation rate distribution appears to behave as
$ \lambda_q \simeq A(1/N)^x $ for $q=1$ and
$ \lambda_q \simeq A'(q/N)^{x'} $ for $q=2$ and $3$,
where
$x \simeq 2.61$, $x' \simeq 2.02$ and $A > A'$.
The wave number $q$ of
the slowest relaxation rate $\lambda_q$ for each $N$ 
is given by $q=2$ for small values of $N$,
while it is given by $q=1$ for large values of $N$.
This crossover corresponds to
the change of the structure of the ring polymer
caused by the localization of the knotted part
to a part of the ring polymer.
}
\kword{
	knot, 
	ring polymer, 
	single polymer, 
	relaxation modes, 
	relaxation rates, 
	Brownian dynamics simulations, 
	topological effects
}
\begin{document}
\maketitle
\section{Introduction}%
\label{sec:Introduction}
\par
The effects of topological constraints caused
by the entanglement of polymers 
on the properties of polymer systems
have attracted much attention.%
\cite{
	deGennes,%
	Doi-Edwards,%
	Hagita2002,%
	Hagita2003,%
	Kauffman,%
	Cloizeaux1981,%
	Deutsch1999,%
	Grosberg2000,%
	Deguchi1997,%
	Shimamura2001,%
	Shimamura2002_1,%
	Shimamura2002_2,%
	Matsuda2003,%
	Quake1994,%
	Lai2002,%
	Lai2001,%
	Wasserman1986,%
	Shaw1993,%
	Stasiak1996,%
	Orlandini1988,%
	Janse1991
}
One of the famous theories on the topological effects
is the reptation theory%
\cite{deGennes,Doi-Edwards} 
of the dynamics of concentrated polymer systems,
where polymers are entangled each other.
In this theory, the topological constraints on a polymer chain
caused by the surrounding polymer chains 
are replaced by an effective tube
confining the polymer chain.
The predictions of the theory
with corrections due to the finiteness of polymer length%
\cite{Doi-Edwards}
agree with the experimental results
and have been 
confirmed by simulations.%
\cite{%
Hagita2002,%
Hagita2003
} 
In concentrated polymer systems,
the entanglement of polymers changes dynamically
and the topological constraints change accordingly. 
In contrast,
in the case of a single knotted ring polymer,
which is one of the self-entangled systems,
the topological constraints
are determined by the knot type and do not change with time.
Therefore, 
a single ring polymer system can be considered as an ideal system
for the study of topological effects.
The investigation of the effects of knots in this system
is expected to  provide a basis for
further understanding of the topological effects in polymer systems. 
\par
The interest in the properties of knotted ring polymers
has been growing in recent years.%
\cite{
	Kauffman,%
	Cloizeaux1981,%
	Deutsch1999,%
	Grosberg2000,%
	Deguchi1997,%
	Shimamura2001,%
	Shimamura2002_1,%
	Shimamura2002_2,%
	Matsuda2003,%
	Quake1994,%
	Lai2002,%
	Lai2001,%
	Wasserman1986,%
	Shaw1993,%
	Stasiak1996,%
	Orlandini1988,%
	Janse1991
}
Especially, the topological effects on the static properties
of ring polymers have been well studied.%
\cite{%
	Cloizeaux1981,%
	Deutsch1999,%
	Grosberg2000,%
	Deguchi1997,%
	Shimamura2001,%
	Shimamura2002_1,%
	Shimamura2002_2,%
	Matsuda2003,%
	Orlandini1988,%
	Janse1991
}
It was conjectured by des Cloizeaux that 
topological constraints make ring polymers
with long chains swell.%
\cite{Cloizeaux1981} 
Although this swelling effect
caused by effective repulsion among segments
is supported by a computer simulation%
\cite{%
Deutsch1999%
}
and
a scaling analysis,%
\cite{%
Grosberg2000%
}
there have been studies%
\cite{%
Deguchi1997,%
Shimamura2001,%
Shimamura2002_1,%
Shimamura2002_2,%
Matsuda2003
}
predicting that the effect vanishes in the long chain limit.
In contrast to the static properties,
the topological effects on the dynamics are
less well studied.%
\cite{Quake1994,Lai2002,Lai2001} 
The equilibrium relaxation of single knotted ring polymers
has been studied by measuring
the time autocorrelation function of the radius of gyration
of a knotted ring polymer
and
the long relaxation time that does not appear for the trivial knot
has been found.%
\cite{Quake1994}
The relaxation time
has the dependence on the number of the essential crossings $C$.%
\cite{Lai2002} 
The nonequilibrium relaxation after cutting 
one bond of a knotted ring polymer
has been studied and 
the distribution of
the nonequilibrium relaxation time of the radius of gyration
has been found to
show different behavior for different knot groups.%
\cite{Kauffman,Lai2001} 
Experimentally, 
knotted circular DNA molecules have been studied by using
gel electrophoresis, electron microscopy, and so on.%
\cite{Wasserman1986,Shaw1993, Stasiak1996} 
\par
In the case of linear polymers,
the relaxation phenomena have been studied systematically 
in terms of the relaxation modes and rates.%
\cite{
	Takano1995,%
	Koseki1997,%
	Hirao1997,%
	Hagita1999-1,%
	Hagita1999-2,%
	Hagita2001,%
	Hagita2002
}
The relaxation modes and rates are given 
as left eigenfunctions and eigenvalues of the time-evolution 
operator of the master equation of the system, respectively.%
\cite{Takano1995,Koseki1997,Hirao1997}
The equilibrium time correlation functions
of the relaxation modes satisfy 
$
	\langle X_p(t) X_q(0) \rangle
	\propto \delta_{p,q} \exp(-\lambda_p t),
$
where 
$X_p$ and $\lambda_p$ denote the $p$th relaxation mode and
its relaxation rate, respectively.
For single linear polymers
represented by the Rouse model, which has
no excluded volume interaction and no hydrodynamic interaction,
the relaxation modes are identical to the Rouse modes,
which have been playing an important role in the theory of
the polymer dynamics.%
\cite{Rouse1953,Zimm1956,Doi-Edwards}
For single linear polymers
with the excluded volume interaction
and without the hydrodynamic interaction,
the relaxation modes and rates are estimated 
by solving a generalized eigenvalue problem
$\sum_{j=1}^{N} C_{i,j}(t_0+\tau)f_{p,j}
= \exp(-\lambda_p \tau)\sum_{j=1}^{N} C_{i,j}(t_0)f_{p,j}$
under the orthonormal condition
$\sum_{i=1}^N \sum_{j=1}^N f_{p,i} C_{i,j}(t_0) f_{q,j}
=\delta_{p,q}$,%
\cite{Koseki1997,Hirao1997}
where
$
	C_{i,j}(t)
$
denotes the equilibrium time-displaced correlation function
$
	\frac{1}{3}
	\langle  \mib{R}_i (t) \cdot \mib{R}_j (0) \rangle
$
of the position of the $i$th segment
relative to the center of mass of the polymer
$\mib{R}_i$
and that of the $j$th segment
and $N$ denotes the number of the segments of the polymer.
The $p$th relaxation mode and the corresponding relaxation rate
are given by $f_{p,i}$ and $\lambda_p$, respectively.
The behavior of the contribution $g_{i,p}$ of
the $p$th slowest relaxation mode to $\mib{R}_i$ is similar to
that of the Rouse mode,
$ g_{i,p} \propto \cos\left[\pi p ( i - \frac{1}{2}) /N \right]$,
and
the corresponding relaxation rate $\lambda_p$ behaves as
$\lambda_p \propto (p/N)^{2\nu+1}$.%
\cite{Doi-Edwards,Koseki1997,Hirao1997}
Here, $\nu \simeq 0.588$ is the exponent for
the power law dependence of the size of a single linear polymer
with the excluded volume interaction
on the number of the segments of the polymer.
\par
The purpose of the present paper is to examine
the effects of the topological constraints on
the relaxation of single ring polymers
by carrying out an analysis similar to that
which has been done for linear single polymers.
Brownian dynamics simulations of 
single ring polymers with the trivial knot and the trefoil knot
are performed,
where the excluded volume interaction is taken into account
and the hydrodynamic interaction is neglected,
and the distribution of relaxation rates is estimated.
\par
In \S \ref{sec:Model},
a model used in the present paper
and the method for the estimation of the relaxation rates
are explained.
The results of the simulations are presented in
\S \ref{sec:Result and Discussion}. 
Summary and discussion are given in the last section.
\section{Model and Relaxation Rates}
\label{sec:Model}
In order to study a single knotted ring polymer in good solvent,
Brownian dynamics simulations of a bead-spring model are performed. 
The dynamics of the $i$th segment of a single ring polymer
with $N$ segments is described by the Langevin equation
\begin{equation}
	\label{eq:Langevin Eq.}
	\frac{ {\rm d} \mib{r}_i(t)}{ {\rm d} t}
	=
	-
	\frac{1}{\zeta}
	\frac{ {\partial} V(\{\mib{r}_j \})}{ {\partial} \mib{r}_i}
	+ \mib{w}_i(t).
\end{equation}
Here, $\mib{r}_i (t)$ denotes the position of the $i$th segment
at time $t$ and $\zeta$ is the friction constant. 
The random force $\zeta \mib{w}_i(t)$ acting on the $i$th segment
is a Gaussian white stochastic process satisfying
$	\langle w_{i,\alpha}(t) \rangle = 0 $ and
\begin{equation}
	\label{eq:random force 1}
	\langle w_{i,\alpha}(t) w_{j,\beta}(t') \rangle
	=
	2
	\frac{k_{{\rm B}}T}{\zeta}
	\delta_{i,j}
	\delta_{\alpha,\beta}\delta(t-t'),
\end{equation}
where $w_{i,\alpha}, k_{{\rm B}}$ and $T$ denote
the $\alpha$-component of $\mib{w}_i$, the Boltzmann constant and
the temperature of the system, respectively. 
The potential $V(\{ \mib{r}_j \}) = V( \mib{r}_1, \cdots, \mib{r}_N)$
describes the interaction between the segments. 
In
eq.\ (\ref{eq:Langevin Eq.}),
the hydrodynamics interaction is neglected.
\par
In the present paper,
we use the potential
given by%
\cite{%
Grest1986,%
Kremer1990,%
Binder,%
Hirao1997%
}
\begin{equation}
	\label{eq:potential}
	V(\{ \mib{r}_j \})
	=  \sum_{i=2}^{N} \sum_{j=1}^{i-1}
			V_{{\rm R}}(|\mib{r}_{i} - \mib{r}_{j}|)
	 + \sum_{i=1}^{N} V_{{\rm A}}(|\mib{r}_{i+1} - \mib{r}_{i}|)
\end{equation}
with
\begin{equation}
	\label{eq:potential LJ}
	V_{{\rm R}}(r)
	=
	\left\{ 
	\begin{array}{lll}
		\displaystyle
			4 \epsilon \Bigl[ 
				 \Bigl( \frac{\sigma}{r} \Bigr)^{12}
				-\Bigl( \frac{\sigma}{r} \Bigr)^{6}
				+\frac{1}{4}
				\Bigr]
			&{\rm for}& r \leq 2^{\frac{1}{6}}\sigma, \\
		0
			&{\rm for}& r >    2^{\frac{1}{6}}\sigma,
	\end{array}
	\right. 
\end{equation}
and
\begin{equation}
\label{eq:potential FENE}
	V_{{\rm A}}(r)
	=
	\left\{ 
	\begin{array}{lll}
		\displaystyle
		-\frac{1}{2}k R_0^2 
		 \ln \Bigl[ 1-\Bigl( \frac{r}{R_0} \Bigr)^2 \Bigr]
				&{\rm for}& r \le R_0, \\
		\infty
				&{\rm for}& r >  R_0.
	\end{array}
	\right. 
\end{equation} 
Here, $V_{{\rm R}}$ is the repulsive part of
the Lennard-Jones potential and represents the excluded volume
interaction between all the segments.
The potential $V_{{\rm A}}$ is called
a finitely extensible nonlinear elastic (FENE) potential
and represents the attractive interaction
between neighboring segments along the ring polymer.%
\cite{Binder}
In the last summation of the right hand side of
eq.\ (\ref{eq:potential}),
$\mib{r}_{N+1} = \mib{r}_1$ because
the $N$th segment is connected to the first segment
along the ring polymer.
\par
As mentioned in the previous section,
the relaxation modes and rates are given
as left eigenfunctions and eigenvalues of the time-evolution 
operator of the master equation of the system, respectively.%
\cite{Takano1995,Koseki1997,Hirao1997}
In the previous studies on the relaxation modes and rates of
single linear polymers,
a trial function for the $p$th relaxation mode is chosen to be
$\mib{X}_p (Q) = \sum_{i=1}^{N}f_{p,i}\mib{R}_i (t_0/2 ; Q)$,
where $Q$ represents a state of the system and
$\mib{R}_i (t_0/2 ; Q)$
denotes the expectation value of $\mib{R}_i$
after a period $t_0 / 2$ starting from a state $Q$.
The quantity $\mib{R}_i$ denotes the 
position of the $i$th segment relative to the center of mass
of the polymer:
$\mib{R}_i = \mib{r}_i - \mib{r}_{\rm c}$
with $\mib{r}_c = \frac{1}{N}\sum_{i=1}^{N} \mib{r}_i$.
For this trial function,
a variational problem equivalent to the eigenvalue problem
for the time-evolution operator leads to
a generalized eigenvalue problem
$
	\sum_{j=1}^{N} C_{i,j}(t_0+\tau)f_{p,j}
	= \exp(-\lambda_p \tau)\sum_{j=1}^{N} C_{i,j}(t_0)f_{p,j}
$
mentioned in the previous section,
where
$
	C_{i,j}(t)
	=
	\frac{1}{3}
	\langle  \mib{R}_i (t) \cdot \mib{R}_j (0) \rangle
$
is the equilibrium time-displaced correlation function.
This analysis is considered to extract
the slow relaxation modes contained in 
the quantities $\{ \mib{R}_i \}$
and to give better results as $t_0$ becomes larger,
because
the contribution of
faster relaxation modes contained in $\mib{R}_i(t_0/2;Q)$
becomes smaller.
\par
In the case of single ring polymers,
the eigenfunctions $f_{p,j}$ of
the above-mentioned generalized eigenvalue problem
are known
because of
the translational invariance
of the segment number along the ring polymer:
$ C_{i,j}(t)  = C_{i+l, j+l}(t)$,
where the subscripts representing the segment numbers
are considered modulo $N$.
The eigenfunctions are given by
$
	f_{q,j} \propto \exp \left( {\rm i} 2\pi qj/N\right)
$
with $q = 1, \cdots, N-1$.%
\cite{Bloomfield1966} 
Note that
the wave number $q=0$ is not included
for it gives a meaningless relaxation mode
$\sum_{j=1}^{N}\mib{R}_j(t_0/2;Q) = \mib{0}$
with an eigenvalue $\exp(-\lambda_q \tau) = 0$,
because
$\sum_{j=1}^{N}\mib{R}_j = \mib{0}$
and
$\sum_{j=1}^{N}C_{i,j}(t) =0$.
For the eigenfunction with the wave number $q$,
the generalized eigenvalue problem is reduced to
\begin{equation}
	\label{eq:eigenvalue problem}
	\hat{C}_q(t_0+\tau) = \exp(-\lambda_q \tau) \hat{C}_q(t_0),
\end{equation}
where
$\hat{C}_q(t)$ denotes the Fourier transform of the correlation matrix
given by
\begin{eqnarray}
	\label{eq:correlation matrix}
	\hat{C}_q(t)
	&=& \sum_{l=0}^{N-1} 
	  C_{i,i+l}(t) \exp \left( {\rm i}\frac{2\pi}{N}ql\right) \\
	\label{eq:correlation matrix 2nd}
	&=& \sum_{l=0}^{N-1} 
	  C_{i,i+l}(t) \cos \left(        \frac{2\pi}{N}ql\right).
\end{eqnarray}
Note that $C_{i,i+l}(t)$ does not depend on $i$
and is identical to $C_{i,i-l}(t)$.
The equilibrium time-displaced correlation function of
the Fourier transform $\hat{\mib{R}}_q$ of $\mib{R}_i$,
which is defined by
\begin{equation}
	\hat{\mib{R}}_q =
	\frac{1}{\sqrt{N}}\sum_{j=1}^{N}
	\mib{R}_j
	\exp \left( {\rm i}\frac{2\pi}{N}qj\right),
\end{equation}
is also given by $\hat{C}_q(t)$ as
\begin{equation}
	\hat{C}_q(t) = \frac{1}{3}
	\langle \hat{\mib{R}}_q (t) \cdot
	\hat{\mib{R}}_{-q} (0) \rangle.
\end{equation}
In the following, we only consider $\hat{C}_q(t)$ with
$
q=1, 2, \cdots, \left\lfloor N/2 \right\rfloor
$,
since $\hat{C}_q(t) = \hat{C}_{N-q}(t)$
as can be seen from eq.\ (\ref{eq:correlation matrix 2nd}).
Here, $\left\lfloor x \right\rfloor$ denotes the floor function 
of a real number $x$.
Equation (\ref{eq:eigenvalue problem}) is equivalent to
the estimation of the relaxation rate $\lambda_q$ 
for the wave number $q$
from the values of $\hat{C}_q(t)$
at times $t_0$ and $t_0 + \tau$
by assuming a single exponential decay of $\hat{C}_q(t)$
at these times.
In fact,
$\hat{C}_q(t)$, which is the autocorrelation function of
the Fourier transform of $\mib{R}_i$,
can be expressed as
$
\hat{C}_q(t) =
a_q^{(1)} \exp ( - \lambda_q^{(1)} t )
+
a_q^{(2)} \exp ( - \lambda_q^{(2)} t )
+
\cdots
$,
where
$0 < \lambda_q^{(1)} <\lambda_q^{(2)} < \cdots$
and $a_q^{(k)} > 0$ for $k = 1, 2, \cdots$.%
\cite{Takano1995,Koseki1997}
Thus,
the estimation of
the slowest relaxation rate
$\lambda_q^{(1)}$ by eq. (\ref{eq:eigenvalue problem})
becomes better as $t_0$ becomes larger,
as mentioned before.
In this paper,
we estimate $\lambda_q^{(1)}$,
the slowest relaxation rate in $\hat{C}_q(t)$
for the wave number $q$,
by fitting
several tens of values of $\hat{C}_q(t)$ at relatively long times
estimated from the simulations to a double exponential decay form
\begin{equation}
\hat{C}_q(t) =
a_q^{(1)} \exp ( - \lambda_q^{(1)} t )
+
a_q^{(2)} \exp ( - \lambda_q^{(2)} t ),
\label{eq:double exponential}
\end{equation}
because this method is more tolerant of statistical errors
in the estimated values of $\hat{C}_q(t)$ than
using eq.\ (\ref{eq:eigenvalue problem}).
\par
In the linearization approximation of Doi and Edwards,%
\cite{Doi-Edwards,Koseki1997}
the relaxation rate $\lambda_p$ of the $p$th Rouse mode
$
\mib{X}_p
= \left( 2/N \right)^{1/2} \sum_{i=1}^{N}
\mib{R}_i \cos \left[
p \pi (i - \frac{1}{2} ) /N
\right]
$
of a single linear polymer
with the excluded volume interaction
is estimated from the static correlation 
$
\langle \mib{X}_p \cdot \mib{X}_p \rangle
$
as
$
\lambda_p^{({\rm LA})} = \left( k_{\rm B}T / \zeta \right)
\left(
\frac{1}{3}
\langle \mib{X}_p \cdot \mib{X}_p \rangle
\right)^{-1}
$.
In the case of single ring polymers,
the relaxation rate $\lambda_q$ for the wave number $q$
is given by the linearization approximation as
\begin{equation}
	\label{eq:linearization approximation}
	\lambda_q^{{\rm (LA)}}
	=
	\frac{k_{{\rm B}}T}{\zeta}
	\left(\frac{1}{3}
	\langle \hat{\mib{R}}_q \cdot \hat{\mib{R}}_{-q} \rangle
	\right)^{-1}
	=
	\frac{k_{{\rm B}}T}{\zeta} \hat{C}_q(0)^{-1}.
\end{equation}
These formulae are derived as follows.%
\cite{Doi-Edwards}
In the linearization approximation,
the equation of motion of each mode $x$
is approximated by the linear Langevin equation
$ \dot{x} = - \lambda x + w $,
where the Gaussian white random force $w(t)$
satisfies $\langle w(t) \rangle =0$ and 
$\langle w (t) w(t') \rangle = 2D \delta(t-t')$.
On the basis of the relation $\langle x^2 \rangle = D / \lambda$,
which holds for this Langevin equation,
the relaxation rate $\lambda$ is estimated
from the mean square amplitude
of the mode $\langle x^2 \rangle$ as
$\lambda = D / \langle x^2 \rangle$.
Thus,
the relaxation rate given by the linearization approximation
becomes small
as the mean square amplitude of the mode vector,
$\mib{X}_p$ or $\hat{\mib{R}}_q$,
becomes large.
The mode vector
$\mib{X}_p$ 
for the linear polymer
roughly corresponds to the end-to-end vector of
a partial chain with a length of $N/p$ segments
and
$\hat{\mib{R}}_q$ 
for the ring polymer
roughly corresponds to the end-to-end vector of
a partial chain with a length of $N/(2q)$ segments.
For a single linear polymer,
the results of the linearization approximation are found to
be consistent with those of
the Monte Carlo simulations
\cite{Koseki1997}
and
the molecular dynamics simulations.
\cite{Hirao1997}
In this paper, 
the relaxation rates $\lambda_q^{({\rm LA})}$
given by the linearization approximation for a single ring polymer
are estimated by using eq.\ (\ref{eq:linearization approximation})
and compared with the relaxation rates $\lambda_q^{(1)}$.
\section{Results of Simulations}
\label{sec:Result and Discussion}
Brownian dynamics simulations of the model described in
the previous section
are performed
for $N=24$, $36$, $48$, $72$, $96$, $144$ and $192$.
The following parameters%
\cite{Hirao1997,Kremer1990}
are used:
$k_{{\rm B}}T/\epsilon=1$, $k\sigma^2/\epsilon = 30$
and $R_{{\rm 0}}/\sigma = 1.5$. 
The Euler algorithm with a time step
${\mit \Delta} t = 1.4 \times 10^{-4}\gamma \sigma^2/\epsilon$
is employed for a numerical
integration of the equation of motion (\ref{eq:Langevin Eq.}). 
Hereafter, we set $\sigma=1$, $\gamma=1$ and $\epsilon=1$. 
\par
The correlation function $\hat{C}_q(t)$ is calculated
from
$
  C_{i,j}(t)
	=
	\frac{1}{3}
	\langle  \mib{R}_i (t) \cdot \mib{R}_j (0) \rangle
$
estimated by the simulations
by using
Eq.\ (\ref{eq:correlation matrix 2nd}). 
The equilibrium average
$
  \langle \mib{R}_{i}(t) \cdot \mib{R}_{j}(0) \rangle
$
is estimated as the time average over $M_{{\rm c}}$ configurations
after the initial $M_{{\rm i}}$ configurations,
which are discarded for the equilibration:
\begin{equation}
	\langle \mib{R}_{i}(t) \cdot \mib{R}_{j}(0) \rangle
	=\frac{1}{M_{{\rm c}}}
	 \sum_{n=M_{{\rm i}}+1}^{M_{{\rm i}}+M_{{\rm c}}}
	 \mib{R}_{i}(n {\mit \Delta} T) \cdot
	 \mib{R}_{j}(n {\mit \Delta} T -t),
\end{equation}
where the configurations are taken from a simulation
at intervals of time ${\mit \Delta} T$.
In the present simulations,
$M_{\rm i}$, $M_{\rm c}$ and ${\mit \Delta} T$ are chosen as
$M_{{\rm i}}{\mit \Delta} T = 10 \tau$,
$M_{{\rm c}}{\mit \Delta} T = 1000 \tau$ and
${\mit \Delta} T = 2 \times 10^{-3} \tau$,
where $\tau$ is
about the longest relaxation time $1/\lambda_{q=1}^{(1)}$
of the ring polymer with the trivial knot for each $N$.
\par
The relaxation rate $\lambda_q^{(1)}$ are
estimated by the weighted least square fit of
$\hat{C}_q(t)$
to eq.\ 
(\ref{eq:double exponential}).
The range of $t$ used for each fit satisfies the conditions
$0.25 < \hat{C}_q(t)/\hat{C}_q(0) < 0.95$ 
and
$\delta \hat{C}_q(t)/\hat{C}_q(t) < 0.1$,
where $\delta \hat{C}_q(t)$ denotes
the statistical error of the estimate of $\hat{C}_q(t)$.
\par
\begin{figure}[!tb]
\begin{center}
		\includegraphics[width=8cm]{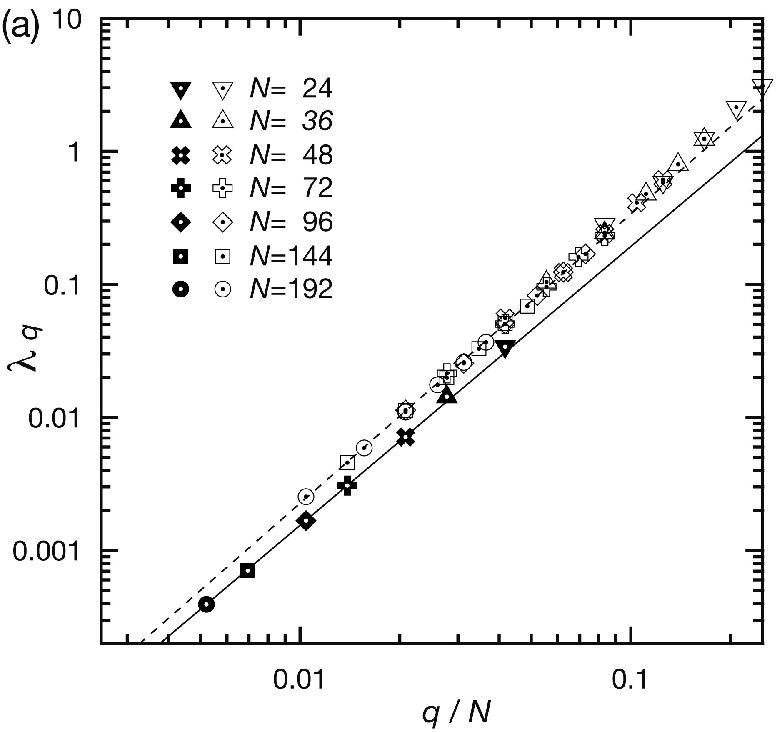} 
	\medskip\\
		\includegraphics[width=8cm]{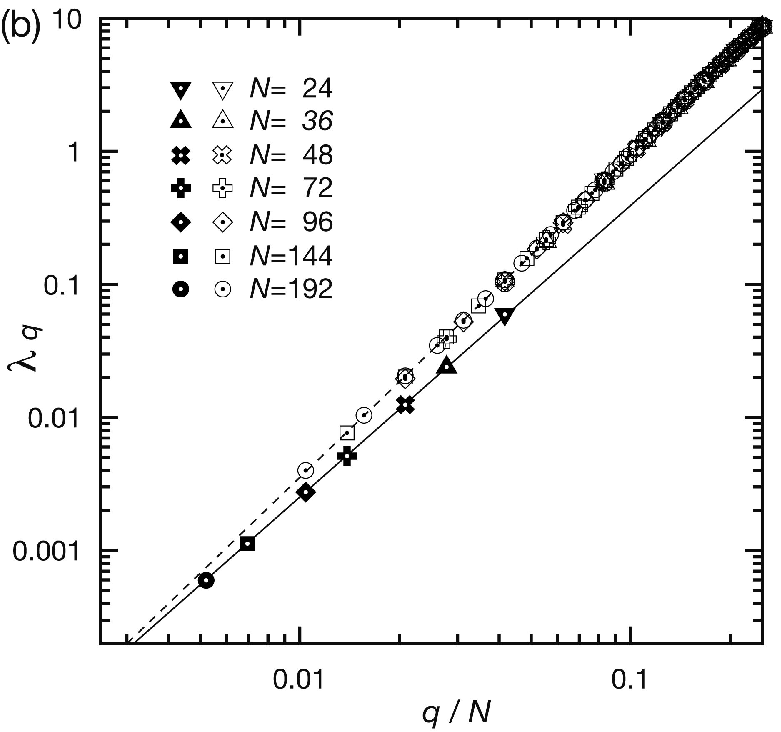} 
	\caption{	
	Log-log plots of
	(a) $\lambda^{(1)}_q$ versus $q/N$
	and
	(b) $\lambda^{\rm (LA)}_q$ versus $q/N$
	for the single ring polymer with the trivial knot,
	where 
	$N=24$, $36$, $48$, $72$, $96$, $144$ and $192$. 
	For each value of $N$,
	the solid symbols represent the relaxation rates for $q=1$
	and the open symbols represent those for $q>1$. 
	In each figure,
	the solid line represents the result of the least square fit
	of the data points for  $q=1$ with $N = 48$--$192$
	to a straight line and
	the broken line represents that
	for  $q > 1$ with $q/N \le 0.05$.
	}
	\label{fig:lambda__K=0}
\end{center}
\end{figure}

Figures \ref{fig:lambda__K=0}(a) and
\ref{fig:lambda__K=0}(b)
show
log-log plots of
$\lambda^{(1)}_q$ versus $q/N$
and
$\lambda^{\rm (LA)}_q$ versus $q/N$,
respectively,
for the ring polymer with the trivial knot,
where
the number of segments
$N=24$, $36$, $48$, $72$, $96$, $144$ and $192$.
The behaviors of
$\lambda^{\rm (LA)}_q$
in Fig.\ \ref{fig:lambda__K=0}(b)
are qualitatively similar to those of $\lambda^{(1)}_q$
in  Fig.\ \ref{fig:lambda__K=0}(a),
although
each value of
$\lambda^{\rm (LA)}_q$ 
does not quantitatively agree with
the corresponding value of
$\lambda^{(1)}_q$,
in contrast to the case of a single linear polymer.%
\cite{Koseki1997,Hirao1997}
In each of Figs.\ \ref{fig:lambda__K=0}(a) and \ref{fig:lambda__K=0}(b),
the data points for $q=1$
and those for $q > 1$
seem to fall on two different straight lines
at small values of $q/N$.
This suggests the power law behaviors
\begin{equation}
	\label{eq:power law}
	\lambda^{(*)}_{q} \simeq 
	\left\{
		\begin{array}{lcl}
		\displaystyle
			A^{(*)}_1 \left(\frac{1}{N}\right)^{x^{(*)}_1}
		&
			{\rm for}
		&
			q=1,
		\bigskip\\
		\displaystyle
			A^{(*)}_{>1} \left(\frac{q}{N}\right)^{x^{(*)}_{>1}}
		&
			{\rm for}
		&
			q>1,
		\end{array}
	\right.
\end{equation}
where the superscript $(*)$ denotes $(1)$ or ${\rm (LA)}$.
The amplitudes and exponents in eq.\ (\ref{eq:power law}) are
estimated by
the least square fit of the data points
to a straight line in the log-log plot:
$A^{(*)}_1$ and $x^{(*)}_1$ are estimated from
the data points
for $q=1$ with $N = 48$--$192$;
$A^{(*)}_{>1}$ and $x^{(*)}_{>1}$ are estimated from
the data points for $q >1$ with $q/N \le 0.05$.
The estimated parameters
for the relaxation rates $\lambda^{(1)}_q$
are given by
$A^{(1)}_1 \simeq 24.0$, $x^{(1)}_1 \simeq 2.10$,
$A^{(1)}_{>1} \simeq 50.2$ and $x^{(1)}_{>1} \simeq 2.17$.
The exponents $x^{(1)}_1$ and $x^{(1)}_{>1}$ are similar to
that for a linear polymer chain
$2\nu+1 \simeq 2.18$.%
\cite{Doi-Edwards,Koseki1997,Hirao1997} 
The separation of the power law dependences
for $q=1$ and $q > 1$
is attributed to
the topological constraints,
because
no such separation appears
for
a single linear polymer%
\cite{Koseki1997,Hirao1997}
and
an ideal single ring polymer,%
\cite{Bloomfield1966}
which have no  excluded volume interaction.
A similar separation of the power law dependences of
the relaxation rates has been observed for
a single linear polymer trapped in an array of obstacles
in two dimensions.%
\cite{Hagita2001}
The parameters for $\lambda^{\rm (LA)}_q$ are estimated as
$A^{\rm (LA)}_1 \simeq 60.9$, $x^{\rm (LA)}_1 \simeq 2.19$,
$A^{\rm (LA)}_{>1} \simeq 214$ and $x^{\rm (LA)}_{>1} \simeq 2.39$.
Note that
the relations $x^{(*)}_1 < x^{(*)}_{>1}$ and $A^{(*)}_1 < A^{(*)}_{>1}$
hold for both 
$\lambda^{(1)}_q$
and
$\lambda^{\rm (LA)}_q$.
It follows from the relations that
the two straight lines in the log-log plot,
which correspond to the power law behaviors
for $q=1$ and $q>1$,
intersects at a small value of $q/N$.
This suggests that
the two power law behaviors for $q=1$ and $q>1$ are only apparent and
eventually merge into a single power law behavior
$\lambda^{(*)}_q \propto (q/N)^X$ with $X = 2 \nu + 1 $
at sufficiently small values of $q/N$.
\begin{figure}[!tb]
\begin{center}
		\includegraphics[width=8cm]{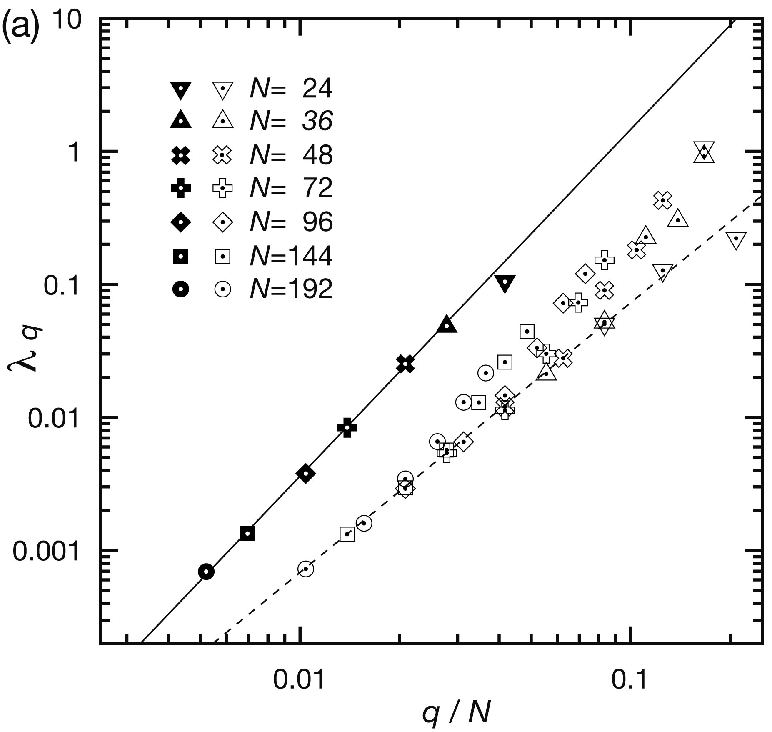} 
	\medskip\\
		\includegraphics[width=8cm]{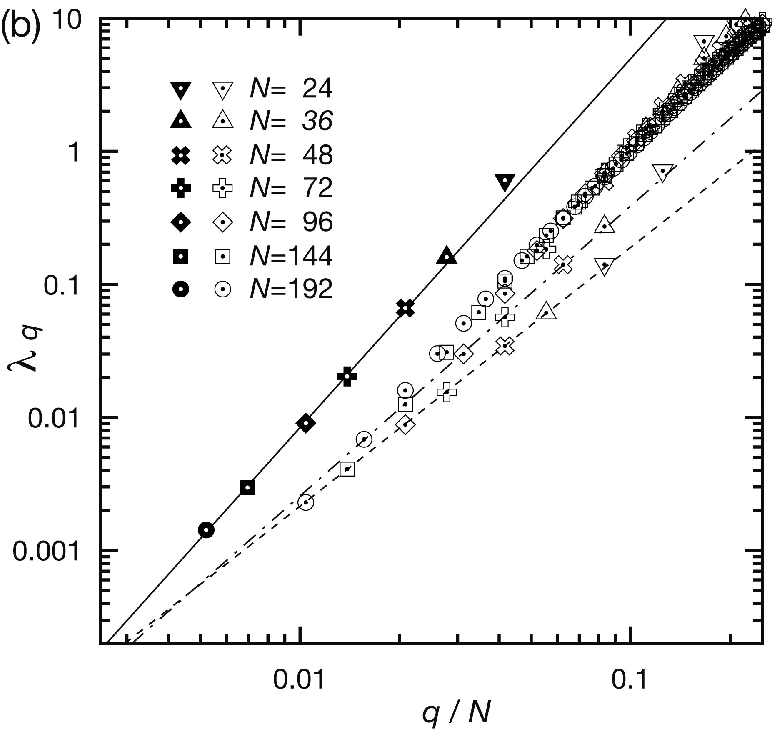} 
	\caption{	
	Log-log plots of
	(a) $\lambda^{(1)}_q$ versus $q/N$
	and
	(b) $\lambda^{\rm (LA)}_q$ versus $q/N$
	for the single ring polymer with the trefoil knot,
	where 
	$N=24$, $36$, $48$, $72$, $96$, $144$ and $192$. 
	For each value of $N$,
	the solid symbols represent the relaxation rates for $q=1$
	and the open symbols represent those for $q>1$. 
	In each figure,
	the solid line represents
	the result of the least square fit of data points for $q=1$.
	The broken line represents that
	for $q=2$ and  $3$ in (a) and that for $q=2$ in (b).
	The dash-dotted line in (b) represents that for $q=3$.
	The data points for $N=48$--$192$ are used for each fit.
	}
	\label{fig:lambda__K=3_1}
\end{center}
\end{figure}
\par
In Figures \ref{fig:lambda__K=3_1}(a) and \ref{fig:lambda__K=3_1}(b),
the $q/N$-dependences of 
$\lambda^{(1)}_q$ and $\lambda^{\rm (LA)}_q$ 
are shown, respectively,
for the ring polymer with the trefoil knot
by a log-log plot,
where
the number of segments
$N=24$, $36$, $48$, $72$, $96$, $144$ and $192$.
In the same way as in Fig.\ 1,
although
values of
$\lambda^{\rm (LA)}_q$ 
in Fig.\ \ref{fig:lambda__K=3_1}(b)
do not quantitatively agree with
those of
$\lambda^{(1)}_q$
in Fig.\ \ref{fig:lambda__K=3_1}(a),
the behaviors of
$\lambda^{\rm (LA)}_q$
are qualitatively similar to those of $\lambda^{(1)}_q$.
In Fig.\ \ref{fig:lambda__K=3_1}(a),
the data points for $q=1$ and those for $q=2$ and $3$
seem to fall on two different straight lines
for small values of $q/N$,
while
the data points for $q=1$, those for $q=2$ and
those for $q=3$
seem to fall on three different straight lines
in Fig.\ \ref{fig:lambda__K=3_1}(b).
Therefore, we assume the power law behaviors
\begin{equation}
	\label{eq:power law 2}
	\lambda^{(*)}_{q} \simeq 
		\begin{array}{lcl}
		\displaystyle
			A^{(*)}_q \left(\frac{q}{N}\right)^{x^{(*)}_q}
		&
			{\rm for}
		&
			q= \mbox{$1$, $2$ and $3$}
		\end{array}
\end{equation}
and estimate $A^{(*)}_q$ and $x^{(*)}_q$
by the least square fit of the data points for each value of $q$
with $N = 48$--$192$ to a straight line in the log-log plot.
The parameters for $\lambda^{(1)}_q$ are estimated as
$A^{(1)}_1 \simeq 591$,
$x^{(1)}_1 \simeq 2.61$,
$A^{(1)}_2 \simeq 7.54$,
$x^{(1)}_2 \simeq 2.02$,
$A^{(1)}_3 \simeq 7.73$ and
$x^{(1)}_3 \simeq 2.04$.
As expected,
the relations
$A^{(1)}_2 \simeq A^{(1)}_3$
and
$x^{(1)}_2 \simeq x^{(1)}_3$ hold,
which suggests that
the relaxation rates
$\lambda^{(1)}_q$ for $q=2$ and $3$  follow
the same power law
\begin{equation}
	\label{eq:power law 3}
	\lambda^{(1)}_{q} \simeq 
		\begin{array}{lcl}
		\displaystyle
			A^{(1)}_{2,3} \left(\frac{q}{N}\right)^{x^{(1)}_{2,3}}
		&
			{\rm for}
		&
			q= \mbox{$2$ and $3$}.
		\end{array}
\end{equation}
The least square fit of the data points for $q=2$ and $3$ 
with $N = 48$--$192$ gives
$A^{(1)}_{2,3} \simeq 7.38$ and
$x^{(1)}_{2,3} \simeq 2.02$.
The parameters for $\lambda^{\rm (LA)}_q$ are estimated as
$A^{\rm (LA)}_1 \simeq 2.97 \times 10^3$,
$x^{\rm (LA)}_1 \simeq 2.77$,
$A^{\rm (LA)}_2 \simeq 16.8$,
$x^{\rm (LA)}_2 \simeq 1.95$,
$A^{\rm (LA)}_3 \simeq 58.8$ and
$x^{\rm (LA)}_3 \simeq 2.18$.
The estimated values of 
$A^{\rm (LA)}_2$,
$x^{\rm (LA)}_2$,
$A^{\rm (LA)}_3$ and
$x^{\rm (LA)}_3$
lead to the intersection of the two straight lines for $q=2$ and $3$
at $(q/N) \simeq 4.3 \times 10^{-3}$,
which suggests that
the two apparent power law behaviors merge into
a single power law behavior
at small values of $q/N$
in the same way as
$\lambda^{(1)}_q$ with $q=2$ and $3$ follows the power law
(\ref{eq:power law 3}).
In Fig.\ \ref{fig:lambda__K=3_1}(a),
the data points of $\lambda^{(1)}_q$ for
$q=4$ with $N=96, 144$ and $192$ seem to fall on
the same straight line
as the data points for $q=2$ and $3$ fall on.
This suggests that
the power law behavior like eq.\ (\ref{eq:power law 3})
is obeyed by
not only for $q=2$ and $3$ but for all values of $q > 1$
at sufficiently small values of $q/N$.
Moreover,
the parameters
$A^{(1)}_{1}$,
$x^{(1)}_{1}$,
$A^{(1)}_{2,3}$
and
$x^{(1)}_{2,3}$,
which satisfy
$x^{(1)}_{2,3} < x^{(1)}_{1}$  and
$A^{(1)}_{2,3} < A^{(1)}_{1}$,
lead to the intersection of
the two straight lines
corresponding to
the power law dependences for $q=1$ and $q = 2$ and $3$
at $q/N \simeq 5.7 \times10^{-4}$.
This suggests that
the two apparent power law behaviors for $q=1$ and $q > 1$
merge into a single power law behavior
at small values of $q/N$.
\par
\begin{figure}[!b]
\begin{center}
	\includegraphics[width=8cm]{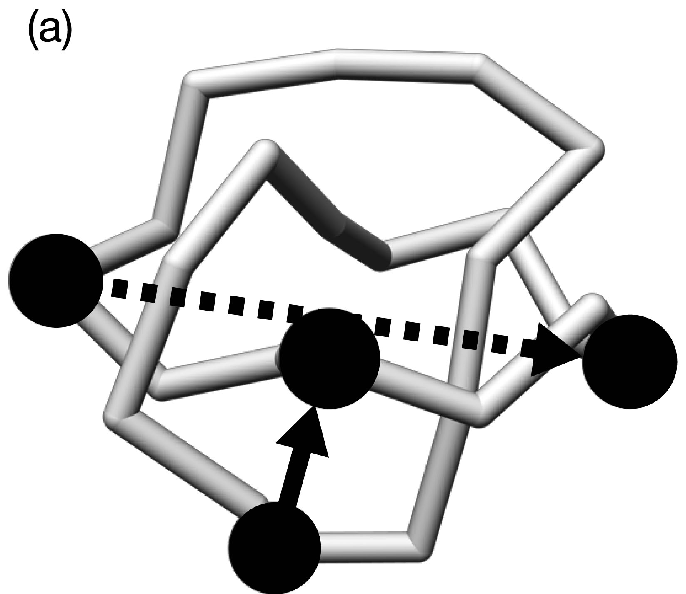}
	\includegraphics[width=8cm]{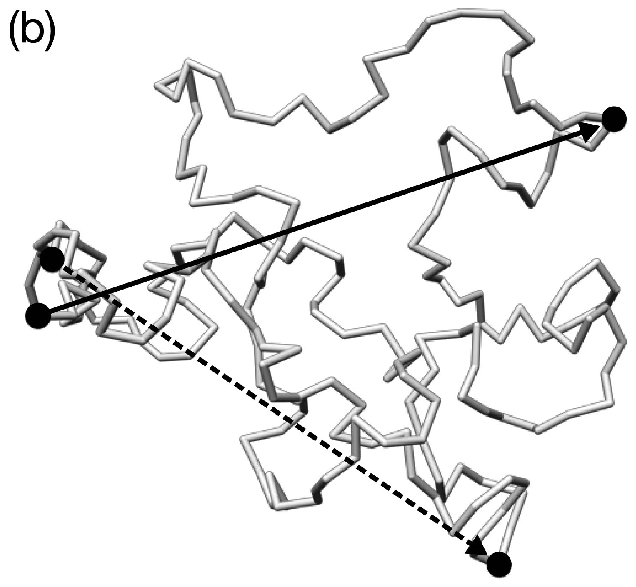}
	\caption{
	Snapshots of the configurations
	of the ring polymer with the trefoil knot
	for (a) $N=24$ and (b) $N=192$.
	The longest end-to-end vector of a partial chain
	with $N/2$ segments and that with $N/4$ segments
	are shown as solid and broken arrows, respectively.
	In each figure,
	the plane of perspective projection is parallel to
	the first and second principal axes of
	the moment of inertia tensor for the four end segments of
	the two end-to-end vectors.
	}
	\label{fig:Snapshot__K=3_1}
\end{center}
\end{figure}

As in the case of the trivial knot,
the separation of the power law dependences of $\lambda^{(1)}_q$
for $q=1$ and $q = 2$ and $3$ appears
for the trefoil knot and
the separation is attributed to the topological constraints.
The pattern of the separation for the trefoil knot is, however,
different from that for the trivial knot.
In the case of the trefoil knot,
$A^{(1)}_{1} > A^{(1)}_{2,3}$ and
$x^{(1)}_{1} > x^{(1)}_{2,3}$ hold,
while
$A^{(1)}_{1} < A^{(1)}_{>1}$ and
$x^{(1)}_{1} < x^{(1)}_{>1}$ hold
for the trivial knot:
In Fig.\ \ref{fig:lambda__K=3_1}(a),
the straight line corresponding to the power law dependence
for $q=1$ lies above that for $q=2$ and $3$,
while the straight line for $q=1$  
is located below that for $q >1$ in Fig.\ \ref{fig:lambda__K=0}(a).
The separation of the power law dependences
for the trefoil knot
leads to the following characteristic behavior of
the relaxation rates.
As $N$ increases,
the wave number $q=q_{\rm min}$ which gives the slowest relaxation rate
$\lambda^{(1)}_{q}$ for each $N$ changes
from $q_{\rm min}=2$ for $N \le 96$ to $q_{\rm min}=1$ for $N \ge 144$.
Because $\lambda^{\rm (LA)}_q$
given by eq.\ (\ref{eq:linearization approximation})
shows the same crossover in Fig.\  \ref{fig:lambda__K=3_1}(b),
the crossover in $\lambda^{\rm (1)}_q$
can be considered to correspond to the 
the crossover in
$	\langle \hat{\mib{R}}_q \cdot \hat{\mib{R}}_{-q} \rangle $:
The largest
$	\langle \hat{\mib{R}}_q \cdot \hat{\mib{R}}_{-q} \rangle $
for each $N$ is given by
$q =2$ for small values of $N$
and $q=1$ for large values of $N$.
Because 
$	\langle \hat{\mib{R}}_q \cdot \hat{\mib{R}}_{-q} \rangle $
roughly corresponds to
the mean square of
the end-to-end distance of
a partial chain with $N/(2q)$ segments as mentioned before,
the crossover indicates that
the end-to-end distance of a partial chain with
$N/2$ segments is shorter than that with $N/4$ segments
for small values of $N$ and is longer for large values of $N$.
This change can be seen in Fig.\ \ref{fig:Snapshot__K=3_1},
which shows
the longest end-to-end vector of a partial chain with $N/2$ segments
and that with $N/4$ segments
in snapshots of the configurations of a ring polymer
with the trefoil knot for $N=24$ and $N=192$.
The longest end-to-end vector 
of a partial chain with $N/2$ segments
is shorter than that with $N/4$ segments for $N=24$
and is longer for $N=192$.
In Fig.\ \ref{fig:Snapshot__K=3_1},
the ring polymer with the trefoil knot is in a ``uniform'' state,
where the knotted part is expanded widely along the polymer,
for $N =24$ and is in a ``phase segregated'' state,
where the knotted part is localized to a part of the ring polymer and
the rest of the ring polymer behaves
like a ring polymer with the trivial knot,
for $N=192$.%
\cite{
Orlandini1988,%
Janse1991,%
Grosberg2000
}
Thus, the crossover from
the state with $q_{\rm min} = 2$
to that with $q_{\rm min} = 1$
corresponds to 
the localization of the knotted part
to a part of the ring polymer.
The localization of the knotted part is considered to be driven
by the entropy gained by the unknotted part
which surpasses the entropy lost by the knotted part.
\par
The localization of the knotted part
of the ring polymer with the trefoil knot
suggests that
the proportion of the knotted part to the remainder part,
which behaves like a ring polymer with the trivial knot,
decreases as $N$ increases.
Therefore,
the behavior of the relaxation rates
of the ring polymer with the trefoil knot
is expected to approach that with the trivial knot 
as $N$ increases.
From Figs.\ \ref{fig:lambda__K=0}(a) and \ref{fig:lambda__K=3_1}(a),
it can be seen that
$\lambda^{(1)}_1$ for the trefoil knot
approaches that for the trivial knot from above as $N$ increases.
In contrast,
$\lambda^{(1)}_{q}$ with $q=2$ and $3$ for the trefoil knot
approach
$\lambda^{(1)}_{q}$ with $q>1$ for the trivial knot from below.
As discussed above,
for each of the trivial and the trefoil knots,
the $q/N$-dependence of $\lambda^{(1)}_q$
is expected to show a single power law behavior
at small values of $q/N$.
Therefore,
it is expected that
the $q/N$-dependence of $\lambda^{(1)}_q$
shows the same single power law behavior
at small values of $q/N$
independently of the knot type
for sufficiently large values of $N$.
\section{Summary and Discussion}
\label{sec:Summary and Discussion}
\par
In this paper,
the relaxation rates of a single ring polymer 
with the trivial knot and the trefoil knot 
are studied through Brownian dynamics simulations
for various values of 
the number of the segments of the ring polymer $N$.
In the case of a single ring polymer, 
each relaxation mode is associated with a wave number $q$
because the ring polymer has 
the translational invariance along the polymer chain. 
The slowest relaxation rate $\lambda^{(1)}_q$
for each wave number $q$ is estimated
by the least square fit of $\hat{C}_q(t)$ given by eq.\
(\ref{eq:correlation matrix})
to
the double exponential decay (\ref{eq:double exponential}).
The linearization approximation $\lambda^{\rm (LA)}_q$
to $\lambda^{(1)}_q$
is also estimated from the static correlation $\hat{C}_q(0)$ 
on the basis of
eq.\ (\ref{eq:linearization approximation}).
The behavior of the distribution of $\lambda^{\rm (LA)}_q$
is qualitatively similar to that of $\lambda^{(1)}_q$,
although each value of $\lambda^{\rm (LA)}_q$
is larger than the corresponding value of $\lambda^{(1)}_q$.
Thus, in the case of a single ring polymer,
the linearization approximation is useful
for studying the behavior of the relaxation rate distribution
qualitatively.
\par
For the trivial knot, 
the relaxation rate $\lambda^{(1)}_q$
appears to behave as
$A^{(1)}_1 (1/N)^{x^{(1)}_1}$ for $q=1$ and 
$A^{(1)}_{>1} (q/N)^{x^{(1)}_{>1}}$ for $q>1$
at small values of $q/N$,
where 
$A^{(1)}_1 < A^{(1)}_{>1}$ and
${x^{(1)}_1} < {x^{(1)}_{>1}}$.
The exponents ${x^{(1)}_1}$ and ${x^{(1)}_{>1}}$
are similar to that found for a single linear polymer chain. 
Even in the case of the trivial knot, 
the effect of the topological constraints appears as
the separation of the power law dependences
of $\lambda^{(1)}_1$ on $q/N$ for $q=1$ and $q >1$.
\par
In the case of the trefoil knot,
the topological constraints causes
the separation of the power law dependences
of $\lambda^{(1)}_q$ for $q=1$ and $q = 2$ and $3$
as in the case of the trivial knot.
In this case,
the amplitudes and the exponents of the power law behaviors
$\lambda^{(1)}_1 \simeq A^{(1)}_1 (1/N)^{x^{(1)}_1}$ for $q=1$ and
$\lambda^{(1)}_q \simeq A^{(1)}_{2,3} (q/N)^{x^{(1)}_{2,3}}$
for $q=2$ and $3$
satisfy the relations
$A^{(1)}_1 > A^{(1)}_{2,3}$ and
$x^{(1)}_1 > x^{(1)}_{2,3}$,
which is different from the case of the trivial knot. 
As the consequence of the separation of the power law behaviors,
the wave number $q=q_{\rm min}$, which gives
the slowest relaxation rate $\lambda^{(1)}_q$ for each $N$,
is given by $q_{\rm min} = 2$ for small values of $N$ and
$q_{\rm min} = 1$ for large values of $N$.
This crossover corresponds to
the change of the structure of the ring polymer
from a state for small $N$,
where the knotted part is extended along the ring polymer,
to another state for large $N$,
where the knotted part is localized to a part of the ring polymer
and the rest of the ring polymer behaves like
the ring polymer with the trivial knot.
It is expected
from the localization of the knotted part
and
the estimated parameters of the power law behaviors
that 
the separated power law behaviors of the relaxation rates
for the trivial knot and the trefoil knots eventually 
merge into a single power law behavior
in the limit of $N \to \infty$,
where the effects of the topological constraints vanish.
\par
In this paper,
we only consider the single ring polymers 
with the trivial knot and the trefoil knot,
which are the simplest knots. 
Even for these simple knots,
the effects of the topological constraints
on the relaxation rate distribution are clearly observed.
Therefore,
it is interesting to study the relaxation rates distribution
of single ring polymers with more complicated knots
or ring polymers with links.
The study in this direction is in progress.
The topological effects seem to vanish in the limit that
the number of the segments of the ring polymer goes to infinity
because of the localization of the knotted part.
The study of the localization of the knotted part
of a single ring polymer is also in progress
by using the average structure of the ring polymer
defined self-consistently
as the average structure of all the sampled structures,
which are translated and rotated
to minimize the mean square displacement from the average structure
itself.
\section*{Acknowledgments}
\label{sec:Acknowledgments}
\par
The authors are grateful to Professor T.\ Deguchi and
Dr.\ K.\ Hagita for fruitful discussions.
This work was partially supported by the 21st Century COE Program;
Integrative Mathematical Sciences: Progress in Mathematics
Motivated by Social and Natural Sciences.
\par\noindent

\end{document}